\documentstyle[12pt]{article}
\def\be{\begin{equation}}
\def\ee{\end{equation}}
\def\bea{\begin{eqnarray}}
\def\eea{\end{eqnarray}}

\def\ap#1#2#3   {{\em Ann. Phys. (NY)} {\bf#1} (#2) #3.}
\def\apj#1#2#3  {{\em Astrophys. J.} {\bf#1} (#2) #3.}
\def\apjl#1#2#3 {{\em Astrophys. J. Lett.} {\bf#1} (#2) #3.}
\def\app#1#2#3  {{\em Acta. Phys. Pol.} {\bf#1} (#2) #3.}
\def\ar#1#2#3   {{\em Ann. Rev. Nucl. Part. Sci.} {\bf#1} (#2) #3.}
\def\cpc#1#2#3  {{\em Computer Phys. Comm.} {\bf#1} (#2) #3.}
\def\err#1#2#3  {{\it Erratum} {\bf#1} (#2) #3.}
\def\ib#1#2#3   {{\it ibid.} {\bf#1} (#2) #3.}
\def\jmp#1#2#3  {{\em J. Math. Phys.} {\bf#1} (#2) #3.}
\def\ijmp#1#2#3 {{\em Int. J. Mod. Phys.} {\bf#1} (#2) #3.}
\def\jetp#1#2#3 {{\em JETP Lett.} {\bf#1} (#2) #3.}
\def\jpg#1#2#3  {{\em J. Phys. G.} {\bf#1} (#2) #3.}
\def\mpl#1#2#3  {{\em Mod. Phys. Lett.} {\bf#1} (#2) #3.}
\def\nat#1#2#3  {{\em Nature (London)} {\bf#1} (#2) #3.}
\def\nc#1#2#3   {{\em Nuovo Cim.} {\bf#1} (#2) #3.}
\def\nim#1#2#3  {{\em Nucl. Instr. Meth.} {\bf#1} (#2) #3.}
\def\np#1#2#3   {{\em Nucl. Phys.} {\bf#1} (#2) #3.}
\def\pcps#1#2#3 {{\em Proc. Cam. Phil. Soc.} {\bf#1} (#2) #3.}
\def\pl#1#2#3   {{\em Phys. Lett.} {\bf#1} (#2) #3.}
\def\prep#1#2#3 {{\em Phys. Rep.} {\bf#1} (#2) #3.}
\def\prev#1#2#3 {{\em Phys. Rev.} {\bf#1} (#2) #3.}
\def\prl#1#2#3  {{\em Phys. Rev. Lett.} {\bf#1} (#2) #3.}
\def\prs#1#2#3  {{\em Proc. Roy. Soc.} {\bf#1} (#2) #3.}
\def\ptp#1#2#3  {{\em Prog. Th. Phys.} {\bf#1} (#2) #3.}
\def\ps#1#2#3   {{\em Physica Scripta} {\bf#1} (#2) #3.}
\def\rmp#1#2#3  {{\em Rev. Mod. Phys.} {\bf#1} (#2) #3.}
\def\rpp#1#2#3  {{\em Rep. Prog. Phys.} {\bf#1} (#2) #3.}
\def\sjnp#1#2#3 {{\em Sov. J. Nucl. Phys.} {\bf#1} (#2) #3.}
\def\spj#1#2#3  {{\em Sov. Phys. JEPT} {\bf#1} (#2) #3.}
\def\spu#1#2#3  {{\em Sov. Phys.-Usp.} {\bf#1} (#2) #3.}
\def\zp#1#2#3   {{\em Zeit. Phys.} {\bf#1} (#2) #3.}
\def\q2{ $q^2$ }

\def\f0{ $f_\pm(0)$ }
\def\e+e-{$e^+e^-$}
\begin{document}
\begin{flushright}
  LNF preprint LNF-95/061(P) 
\end{flushright}
\begin{center}
\vspace*{2.0cm}
\LARGE{CHARM LIFETIMES AND SEMILEPTONIC DECAYS } \\
\vspace*{2.0cm} 
\large{Stefano Bianco}\\
\large{{\it Laboratori Nazionali di Frascati dell'INFN \\
 Via E.Fermi 40, I-00044 Frascati ITALY }}
\end{center}
\vspace*{3.0cm}
\begin{abstract}
This paper gives an experimental overview
of current status and open questions on charm lifetimes and semileptonic
decays, the latter including new results presented by E687 and CLEO at this
Conference.
\end{abstract}
\vspace*{1.0cm}
\begin{center}
\end{center}
\vspace*{2.0cm}
\begin{center}
{\it Invited mini-review talk given at the \\
1995 International Europhysics Conference on High Energy Physics\\
   July 27 August 2, Brussels, Belgium. 
}
\end{center}
\vfill\eject
\section{Introduction}
The study of lifetimes and semileptonic (SL) decays of charmed 
mesons and baryons is
currently the goal of several experiments all around the world.
\begin{eqnarray}
\tau = {{\hbar}\over{\Gamma_{leptonic}+\Gamma_{SL}
+\Gamma_{nonleptonic}}}
 \label{eq:width}
\end{eqnarray}
This paper is focussed on giving  an
overview of current status and open questions, and on reviewing 
new results by E687 and CLEO presented as contributed papers at this Conference.
A wealth of detailed reviews exists on both charm
lifetimes\cite{malvezzi95}-\cite{forty94} and SL decays
\cite{wiss94}-\cite{perasso95}-\cite{richman95},
some very up-to-date, 
which the reader can refer to for details.
\section{Lifetimes}
In the case of charmed hadrons, the picture of states being made
of a heavy quark $Q$ and a cloud of light quarks $q$ and gluons,
with essentially a unique lifetime independent of both the number of light
partners and  their flavor, is grossly
contradicted by experiment: the lifetimes differ widely both between 
mesons and baryons $(D^+/\Xi^0_c \sim 10)$, and among mesons 
$(D^+_s/D^0=1.125\pm 0.042, D^+/D^0=2.547\pm 0.043)$. Since 
the SL partial widths are compatible:
\begin{equation} 
B(D^+\rightarrow e^+ X)= (17.2\pm 1.9) \% \,(PDG94)
\label{eq:inclbr+}
\end{equation}
\begin{equation}
 B(D^0\rightarrow e^+ X)= 
(6.97\pm 0.18 \pm0.30) \%  \,(CLEO)
\label{eq:inclbr0}
\end{equation}
\bea 
{{\Gamma (D^0\rightarrow e^+ X)}\over{\Gamma(D^+\rightarrow e^+ X)}} & = &
{{B(D^0\rightarrow e^+ X)}\over{B(D^+\rightarrow e^+ X)}}      
{{\tau(D^+)}\over{\tau(D^0)}}            \\
& = & 1.03\pm.12
\label{eq:pwsl}
\eea
while the leptonic partial widths are negligible:
\begin{equation}
\Gamma_{leptonic}\sim 10^{-3}-10^{-4}                  
\label{eq:pwlep}
\end{equation}
as shown in eq.\ref{eq:width} the extra rate is in the nonleptonic width. 
W-exchange diagrams (partially helicity- and color-suppressed),
W-annihilation diagrams, Pauli interference of the decay and the spectator
quarks, all conspire in increasing the $D^+$ lifetime with respect to the other
charmed mesons. 
Theoretical consensus does exist on the prediction
\begin{equation}
  \tau(D^+) > \tau(D^0) \sim \tau(D^+_s)
\end{equation}
while the extent of the $\sim$ sign is not clear.
In the charmed baryon sector,  on the other hand, the presence
of a second light quark makes W-exchange neither helicity- nor color-suppressed,
thus increasing the spread in lifetimes.
Predictions span quite
widely\cite{guberinavoloshincheng}-\cite{blok}:
\be
 \tau(\Omega_c^0) < \tau(\Xi_c^0) < \tau(\Lambda_c^+) \sim \tau(\Xi_c^+)
\ee
\be
 \tau(\Omega_c^0) \sim \tau(\Xi_c^0) < \tau(\Lambda_c^+) < \tau(\Xi_c^+)
\ee
\be
 \tau(\Omega_c^0) < \tau(\Xi_c^0) \quad {\rm OR} \quad 
 \tau(\Omega_c^0) > \tau(\Xi_c^+)
\ee
Charm lifetimes are
therefore a probe for hadronic dynamics; they also are a tool for $b$-physics
items such as tagging, $B^0-\bar{B}^0$ mixing, and CP violation.
\par
\begin{table}\begin{center}\caption{ Measurements of the $\Omega^0_c$ 
lifetime (ps), beams, and decay channels.}
\vspace{0.3cm}
\begin{tabular}{l|l} \hline\hline
              E687 \cite{e687omegac}  & WA89 \cite{wa89omegac}  \\ \hline
                                       &                         \\
  $0.086^{+0.027}_{-0.020}\pm0.028$ &
  $0.055\,^{+0.013}_{-0.011}\,^{+0.018}_{-0.023}$                 \\
  $\gamma$, 220\,GeV                  &       $\Sigma^-$, 340~GeV/c \\
  $\Sigma^+ (p\pi^0,n\pi^+) K^- K^+ \pi^+$                         &
  $\Xi^-K^-\pi^+\pi^+, \Omega^-\pi^+\pi^-\pi^+$                    \\   
\hline\hline
\end{tabular}
\end{center}
\label{omegac}
\end{table}                                                            
Experimentally, the field is mature for charm mesons. Over the past 10 years
fixed-target photoproduction experiments E691 and, recently, E687 
\cite{e687lifetimes} have provided
precision measurements well beyond the prediction power of theory. They both
employed a microstrip silicon detector for the detection of 
production and decay vertices, and
the measurement of decay lengths.
\par
The scenario for baryons is less statistically satisfactory. Still, a definite
hierarchy has  emerged
\bea
 \tau(\Omega_c^0) \le \tau(\Xi_c^0) < \tau(\Lambda_c^+) < \tau(\Xi_c^+) <
\nonumber \\
<   \tau(D^0)    < \tau(D^+_s)      < \tau(D^+)
\eea
with the E687 measurement of the $\Omega_c$
lifetime \cite{e687omegac} being confirmed at this Conference by the new WA89
value \cite{wa89omegac} 
(Table 1).
%
%
\par
Table \ref{lifetime} shows the most precise measurements compared to the
PDG94 average. 
\begin{table}\begin{center}\caption{ Lifetimes of $D^+$, $D^0$, $D^+_s$,
$\Lambda^+_c$, $\Xi^0_c$, $\Xi^+_c$ (ps)}
\vspace{0.3cm}
\begin{tabular}{c|l|l} \hline\hline
       &        &       \\
       & E687   & PDG94 \\ \hline
       &        &       \\
$D^+$        & $1.048\pm0.015\pm0.011$ & $1.057\pm0.015$ \\  
$D^0$        & $0.413\pm0.004\pm0.003$ & $0.415\pm0.004$ \\  
$D^+_s$      & $0.475\pm0.020\pm0.007$ & $0.467\pm0.017$ \\  
$\Lambda^+_c$& $0.215\pm0.016\pm0.008$           & $0.200^{+0.011}_{-0.010}$\\  
$\Xi^0_c$    & $0.101^{+0.025}_{-0.017}\pm0.005$ & $0.098^{+0.023}_{-0.015}$\\  
$\Xi^+_c$    & $0.41 ^{+0.11}_{-0.08}\pm0.02$ & $0.35^{+0.07}_{-0.04}$\\  
\hline\hline
\end{tabular}
\end{center}
\label{lifetime}
\end{table}                                                            
\par
\section{Semileptonic decays}
Semileptonic decays are relatively simple to handle theoretically,
since the decay matrix decouples into a weak current
(describing the $W\ell\nu_{\ell}$ vertex), and a strong current (for the
$Wc\bar q$ vertex) that  is parameterized through functions
-- form factors (f.f.) -- of the  invariant mass $(q^2)$ of the $W$ exchanged.
Measuring the f.f. and their \q2 evolution provides useful insights into quark
dynamics, model-dependent information on absolute branching ratios, and is a
playground for theories on interpolation to the beauty-related CKM elements
$V_{ub}$ and $V_{cb}$. Experimentally, rates are not too small, but the
undetected neutrino forbids us to close the event kinematically, 
and we are left with demonstrating that the final state is indeed exclusive. 
\par
Experiments studying
SL in \e+e- annihilations exploit the favorable charm-to-background ratio,
while having to cope with the relatively small cross section; they also
usually enjoy excellent $\gamma$ and $\pi^0$ reconstruction capabilities. 
\par
In
fixed-target hadroproduction the charm cross section is higher, with the
background of light quarks also larger. Fixed-target photoproduction is
somehow midway, with more favorable signal-to-noise. Crucial is the
possibility of exploiting excellent primary and secondary vertexing.
\par
Common techniques to both \e+e- and fixed-target are $D^*$-tagging (i.e., 
selecting D's coming from the $D^*\rightarrow D\pi$ decay), wrong-sign
subtraction, kinematic cuts, and particle identification \cite{richman95}.
\subsection{$D\rightarrow$ {\it( Pseudoscalar)} $\ell \nu$}
The differential decay rate 
for the decay of a 
charmed meson to a pseudoscalar meson, a lepton, and a neutrino
has the dependence 
\begin{equation}
\label{decrate}
\frac{d\Gamma}{dq^2} = \frac{G^2_F |V_{cq}|^2 P^3}{24\pi^3} 
\left\{ |f_+(q^2)|^2+ |f_-(q^2)|^2{\cal O}(m_\ell^2)+... \right\}
\end{equation}
where $P$ is the momentum of the pseudoscalar meson in the reference frame of 
the charmed meson, and the $Wc\bar q$ vertex is described by only
two f.f.,  $f_\pm(q^2)$. 
\par
The $f_\pm(q^2)$ form factors [note that $f_-(q^2)$ becomes unimportant in
the limit of zero lepton mass]
are usually parameterized as
\begin{equation}
F_\pm(q^2) = \frac{f_\pm(0)}{(1-q^2/M^2_{pole})}
\label{vecmes}
\end{equation}
or
\begin{equation}
F_\pm(q^2) =  f_\pm(0)e^{\alpha q^2}.
\label{wisgur}
\end{equation}
The form in eq.\ref{vecmes} relies \cite{wirbel85}-\cite{korner90}
on the coupling of 
the $c\bar q$ quarks to the
virtual $W^\pm$ being dominated by bound states of the $c\bar q$ 
system; in the case of  $D\rightarrow K\ell \nu$ decay, one expects that
$M_{pole}$ should be set to the mass of the vector $D^*_S(2110)$ since it
has the same spin-parity  as the $c\bar s$ current. Equation \ref{wisgur}
is suggested
by the  ISGW model \cite{isgur89}. However, the two forms are not
distinguishable in the  range probed by $K\ell\nu$ decays 
$(q^2 < 2\, GeV^2/c^4)$, while sensitivity exists for $\pi\ell\nu$ decays 
$(2< q^2 < 3\, GeV^2/c^4)$ (ref.\cite{wiss94}).
\par
E687 has recently reported\cite{johns94} 
on the analysis of 1897$\pm$62 events in the decay mode 
$D^0\rightarrow K^-\mu^+\nu_\mu$. The strategy followed was to increase the
event statistics by relaxing the $D^*$-tag requirement\cite{johns95}. 
Results are
shown in 
Table 4,
%
%
including the first measurement of the f.f.
ratio at $q^2=0$, which is consistent with the theoretical estimates, which
range from -1.2 to -0.4. The $M_{pole}$ is now $2.2\sigma$ away from the
$D^*_s(2110)$. Finally, the ratio 
\be
\frac{B(D^0\rightarrow K^{*-}\mu^+\nu_\mu)}
     {B(D^0\rightarrow K^-\mu^+\nu_\mu)}
\label{eq:brk*k}
\ee
in agreement with previous measurements and different from unity as so far
predicted by theory, confirms the so-called vector-to-pseudoscalar puzzle.
\par  
The CLEO group has submitted to this Conference \cite{eps95cleo} the 
first measurement
of the branching ratios of the decays $D_s^+\rightarrow \eta\ell^+\nu_\ell$ and 
$D_s^+\rightarrow \eta'\ell^+\nu_\ell$ [where $\ell=\mu (mostly), e$], relative
to $B(D_s^+\rightarrow \phi\ell^+\nu_\ell)$. This measurement constitutes a
real steeplechase for their new CsI calorimeter \cite{cleodet}. The
calorimeter is used for detecting many-photon final states, such as
$\eta\rightarrow \gamma\gamma$, $\eta'\rightarrow \eta (\rightarrow
\gamma\gamma)\pi^+\pi^-$, and also for the tagging
mode $D^{*+}_s\rightarrow D_s^+\gamma$ used to
clean up the high statistics mode. Their first measurement
(Table 3)
%
%
is consistent with the 0.56$\pm$0.06 
world average of the ratio (eq.\ref{eq:brk*k}), and confirm the existence of
a vector-to-pseudoscalar puzzle. 
A recent paper \cite{isgw2vtop} is also quoted, 
in which a prediction of 0.6 is derived for the vector-to-pseudoscalar   
ratio, in good agreement with their result.
\begin{table}
\begin{center}\caption{ 
$B(D_s^+\rightarrow \phi e^+\nu)/B(D_s^+\rightarrow (\eta+\eta') e^+\nu)$
(the ISGW2 predictions are for -10$^o$ and -20$^o$ $\eta-\eta'$ mixing angles
respectively)
}
\vspace{0.3cm}
\begin{tabular}{l|l} \hline\hline
               &       \\
CLEO first measurement \cite{eps95cleo} & $0.60\pm0.06\pm 0.06$ \\ \hline
ISGW2 \cite{isgw2vtop} & 0.60, 0.69     \\
\hline\hline
\end{tabular}
\vspace{0.3cm}
\end{center}
\label{cleoeps}
\end{table}                                                            
\par
\par
The importance of the Cabibbo-suppressed decays 
 $D\rightarrow \pi\ell\nu$ deserves to be stressed, 
since they allow one to probe f.f. models over an extended \q2 range.
CLEO has measured $|f_+^\pi(0)/f_+^K(0)|$ in both 
$D^+\rightarrow \pi^0\ell^+\nu_\ell$ \cite{cleopi0}
and 
$D^0\rightarrow \pi^-\ell^+\nu_\ell$\cite{cleopi+}
 channels. Results are consistent with
theoretical predictions ranging from 0.7 to 1.4. The challenging
$\pi^0$ channel does not suffer from the $\pi^{\pm}/K^{\pm}$ 
misidentification background.
\subsection{$D\rightarrow$ {\it( Vector)} $\ell \nu$}
The decay of a charm meson to a vector meson involves a
hadronic current described by two axial $A_{1,2}(q^2)$ and one vector 
$V(q^2)$ f.f. (neglecting   $ m_\ell^2$ terms). 
Traditionally\cite{pdg94p1566}-\cite{anjos90} one measures the ratios 
$R_V\equiv V(0)/A_1(0)$, $R_2\equiv A_2(0)/A_1(0)$, 
and assumes a pole form for the f.f.'s with
$M_A=M_{D^{**}_s}=2.5\,GeV/c^2$ and $M_V=M_{D^{*}_s}=2.1\,GeV/c^2 $.
\par
The decay $D^+\rightarrow \bar{K}^{*0}\ell^+\nu_\ell$ has been studied by E691,
E653, and E687. No new measurements are available, and the latest results 
on $R_V, R_2$ and 
polarizations are quite consistent and agree with lattice calculations.
\par
The decay $D^+\rightarrow \phi\ell^+\nu_\ell$ is also important, since it can be
used to obtain a  model-dependent estimate of the absolute branching ratio for
$D^+_s\rightarrow \phi \pi^+$:
\bea
B(D^+_s\rightarrow \phi \pi^+) & = & \tau(D_s) \times             \nonumber \\
 model\rightarrow & & {{\Gamma(D^+_s\rightarrow \phi \ell^+ \nu_\ell)}\over
     {\Gamma(D\rightarrow \overline{K}^* \ell^+\nu_\ell)}}\times  \nonumber \\
 & & \Gamma(D\rightarrow \overline{K}^* \ell^+\nu_\ell) \times    \nonumber \\
 & & {{\Gamma(D_s^+\rightarrow \phi \pi^+)}\over
     {\Gamma(D^+_s\rightarrow \phi \ell^+ \nu_\ell)}}                       
\eea
This decay has been studied by E653\cite{kodama93}, E687\cite{frabetti94},
 and CLEO-II\cite{avery94}.
No new results are available, errors are large, and no clear agreement is found
either among experiments, or with the $K^{*0}$ channel.
\subsection{Semileptonic Baryon Decays}
Both CLEO\cite{bergfeld94}-\cite{crawford95}
 and ARGUS\cite{albrecht94} have data on the decay 
$\Lambda_c\rightarrow
\Lambda\ell^+\nu$, while data on 
$\Xi_c\rightarrow \Xi\ell^+\nu$ only come from CLEO\cite{alexander94}.
Particularly interesting is the measurement of the polarization of the
$\Lambda$ in the final state, for which an explicit prediction is made by HQET
\cite{robertsmannelkorner}.
Finally, CLEO evidence for $\Omega_c \rightarrow \Omega e^+\nu$ was 
shown recently \cite{perasso95}, 
new results with improved statistics are expected soon
\cite{menaryprivate}.
\par
\begin{table}
  \caption{Preliminary results for the E687 new measurement of
$D^0\rightarrow K^-\mu^+\nu_\mu$.}
\vspace{0.5cm}
\begin{center}
\begin{tabular}{||c|c|c|l||} \hline\hline
& & & \\
Reference & $\frac{B(D^0\rightarrow K^-\mu^+\nu_\ell)}{B(D^0\rightarrow 
            K^-\pi^+)}$
                        & $M_{pole} (GeV/c^2)$ 
& $\frac{B(D^0\rightarrow K^{*-}\mu^+\nu_\mu)}
        {B(D^0\rightarrow K^-\mu^+\nu_\mu)}$  \\
\hline
\cite{johns94} (E687 95) & 
0.852$\pm$ 0.034$\pm$0.028 & 
1.87$^{+0.11+0.07}_{-0.08-0.06}$ &
 0.62$\pm$0.07$\pm$0.09 \\
\cite{cleo93} (CLEO 93) &
0.978$\pm$ 0.027$\pm$0.044 & 
2.00$\pm$ 0.12$\pm$0.18 & 
0.62$\pm$ 0.08 \\
\cite{cleo91} (CLEO 91) &
0.79$\pm$ 0.08$\pm$0.09 $(\mu\,only)$ &  
2.0$^{+0.4+0.3}_{-0.2-0.2}$ &
0.51$\pm$ 0.18$\pm$0.06    \\
\cite{anjos89}-\cite{anjos91} (E691) &
0.91$\pm$ 0.07$\pm$0.11 &  
2.1$^{+0.4}_{-0.2} \pm 0.2$ &
0.55$\pm$ 0.14    \\
\hline\hline
Reference 
& $\Gamma (D^0\rightarrow K^- \ell^+\nu_\mu) 10^{10} s^{-1}$ 
& $f_+(0)$
& $f_-(0)/f_+(0)$           \\
\hline
\cite{johns94} (E687 95) & 
8.07$\pm$0.37 $\pm$ 0.44 & 
0.71$\pm$0.03$\pm$0.02   &
-1.3$^{+3.6}_{-3.4}\pm{0.6}$ \\
\cite{cleo93} (CLEO 93)  &
9.1$\pm$0.3 $\pm$ 0.6    & 
0.77$\pm$0.01$\pm$0.04   &
                            \\
\cite{cleo91} (CLEO 91)  &
                         & 
0.81$\pm$0.03$\pm$0.06   &
                            \\
\cite{anjos89} (E691) &
9.1$\pm$1.1 $\pm$ 1.4    & 
0.79$\pm$0.05$\pm$0.06   &
                            \\
\hline\hline
\end{tabular}
\vspace{0.5cm}
\end{center}
\label{tab:e6872bod}
\end{table}                                                     
\par                                                    
\section{Conclusions}
Charm lifetimes are a stage for precision physics: E687
has measured meson lifetimes at a level of precision
beyond that presently predictable by theory, while 
measuring the lifetimes of all known charm baryons, including the first
measurement of the $\Omega_c^0$. The  new measurement presented
by WA89 at this Conference confirms that the $\Omega_c^0$ has the shortest
lifetimes of all known charm particles. New data are expected by CERN WA89, 
and Fermilab E791, E831, SELEX .
\par
Precise, consistent data is available for the 
SL decays of charm mesons to pseudoscalar light mesons. New E687
results have been presented on $K\mu\nu$ decay, including the first measurement
of the f.f. ratio at $q^2=0$. Due to the limited \q2 range accessible, one
cannot tell the pole form from the exponential: new data 
in
the Cabibbo-suppressed channel $\pi\ell\nu$ would be welcome,
to extend the \q2 range.
CLEO has presented the first measurement of the $D\rightarrow
(\eta,\eta')\ell\nu$ branching ratios, 
which confirms the so-called vector-to-pseudoscalar
puzzle. 
Data on SL decays to vector light mesons are still confused, 
partially in disagreement with theory, with large errors. New results should be
expected from Fermilab E831, E791 and Beijing BES. 
\setcounter{secnumdepth}{0} 
\section{Acknowledgements}
I would like to thank all my E687 fellow collaborators for their efforts,
continuous discussions and information.
Particular thanks go to G.Bellini, J.Cumalat, S.Malvezzi, L.Perasso and S.Ratti
(E687), S.Paul and  L.Rossi (WA89), S.Menary (CLEO).
I finally thank the Organizers for a perfectly enjoyable Conference.

\end{document}